\documentclass[conference]{IEEEtran}
\ifCLASSINFOpdf
\else
\fi
\usepackage{xfrac}
\usepackage{amsmath}
\usepackage{multirow}
\usepackage{color}
\usepackage{amssymb}
\usepackage{cite}
\DeclareMathOperator*{\argmin}{argmin}
\usepackage{lipsum}
\usepackage{algpseudocode,algorithm}

\ifCLASSOPTIONcompsoc
\usepackage[caption=false,font=normalsize,labelfon
t=sf,textfont=sf]{subfig}
\else
\usepackage[caption=false,font=footnotesize]{subfi
g}
\fi

\usepackage{stfloats}

\begin{document}
\bstctlcite{IEEEexample:BSTcontrol}


\title{Improved Error Performance in NOMA-based Diamond Relaying}
%
%
%

\author{\IEEEauthorblockN{
Ferdi KARA, Hakan KAYA}
\IEEEauthorblockA{
Wireless Communication Technologies Laboratory (WCTLab) \\
Department of Electrical and Electronics Engineering\\
Zonguldak Bulent Ecevit University\\
Zonguldak, TURKEY 67100\\
Email: \{f.kara,hakan.kaya\}@beun.edu.tr}
}

\maketitle
\begin{abstract}
Non-orthogonal multiple access (NOMA)-based cooperative relaying system (CRS) has emerged as a solution to the spectral inefficiency problem of the conventional CRS thanks to the NOMA integration. Thus, as a subset of the NOMA-CRS, the NOMA-based diamond relaying network (NOMA-DRN) also provides a performance gain in terms of throughput. However, the NOMA-DRN has a poor error performance due to the second phase (uplink), indeed, it has an error floor regardless of the transmit power, power allocation and channel qualities. To address this problem, in this paper, we propose a novel NOMA-DRN scheme where a joint maximum likelihood (JML) decoding is implemented at the destination. Then, we define the performance metrics (i.e., bit error rate (BER) and the diversity order) of the NOMA-DRN with the JML and analyze the computational complexity. Moreover, we demonstrate that the new NOMA-DRN with JML can cope with the error floor penalty of the conventional NOMA-DRN. Hence, a spectral efficient NOMA-CRS scheme can be achieved with high data reliability. Specifically, this improvement can reach to $\sim20-30dB$ in the transmit power which is superb gain in terms of energy efficiency perspective. Furthermore, with the proposed NOMA-DRN with the JML, the full diversity order can be achieved in the low-medium SNR region. 
\end{abstract}

\IEEEpeerreviewmaketitle
\section{Introduction}
Cooperative relaying scheme has been one of the most attractive topics since its first applications \cite{Laneman2004}. It provides a spatial diversity when multiple antennas can not be placed due to the physical limitations at the transmitter and/or receiver. Besides, it also makes possible the higher coverage area when the direct link is not available between the source and the destination. Therefore, the cooperative relaying scheme has been indispensable in the wireless communication evolution as well as the standards for almost two decades. However, the cooperative relaying scheme suffers from the spectral inefficiency since it occupies more than one resource block (time slots) for the forwarding strategies. To address this problem, Non-orthogonal multiple access-based cooperative relaying system (NOMA-CRS) has been proposed in the literature \cite{Kim2015a} where two consecutive symbols are conveyed simultaneously thanks to the NOMA technique. In NOMA, multiple symbols are merged with different power allocation (PA) coefficients and transmitted to the destination (users) on the same resource block (time, frequency, code). The interference mitigation is achieved by the successive interference canceler (SIC) at the receivers so that all symbols can be detected \cite{Saito2013}. In the first NOMA-CRS paper \cite{Kim2015a}, it is proved that the spectral efficiency is improved compared to the conventional CRS. Then, the NOMA-CRS schemes have attracted a remarkable attention from the both academia and industry where  NOMA-CRS schemes have been analyzed in terms of different key performance indicators such as sum-rate, outage probability, energy efficiency, bit error probability over various wireless channel models \cite{Jiao2017,Xu2016,Zhang2018,Abbasi2019,Kader2019}. In those works, the theoretical analysis is conducted to reveal the superiority of the NOMA-CRS in terms of capacity and outage probability. However, they mostly assume the perfect SIC and when this assumption is relaxed, the error performance penalty of the NOMA-CRS is presented \cite{Kara2020b}. In addition to NOMA-CRS with a single relay, multiple relay scenarios have been also considered in the literature \cite{Kim2016}. In the multiple relay NOMA-CRS, relay selection schemes have been analyzed in terms of capacity and outage probability. 

Moreover, when two relays are located between the source and the destination, as a subset of NOMA-CRS, called NOMA-based diamond relaying network (DRN) is proposed in \cite{Wan2019} and the sum-rate expression is derived. Then, the optimum power allocation is studied in \cite{9057498} to maximize the sum-rate of the NOMA-DRN. However, when the error probability of the NOMA-DRN is analyzed, it is seen that the NOMA-DRN has a poor error performance \cite{Kara2019}. Although its performance improvement in terms of sum-rate capacity, the NOMA-DRN has an error floor since it includes an uplink phase of NOMA \cite{Kara2018f}. Regardless of the transmit power and the PA, the NOMA-DRN fails due to the SIC at the destination. Nevertheless, the authors in \cite{Yeom2019 } prove that the error performance of the uplink NOMA can be improved by using joint maximum-likelihood (JML) detector rather than the SIC detector. 

In this paper, to this end, we propose a novel NOMA-DRN with the JML to improve the error performance of the NOMA-DRN with SIC receiver. We revealed that the proposed NOMA-DRN with the JML achieves a good error performance and the error floor exists no more. Therefore, the capacity enhancement of the NOMA-DRN can be provided without the error floor penalty so that a reliable communication is achieved. Moreover, for the same error performance target, the proposed NOMA-DRN with JML can save $\sim 20-30dB$ in transmit power which is very promising for the energy-constraint networks such as Internet of Things (IoT) applications. In addition to these advantages, the NOMA-DRN with the JML offers a full diversity order in the low-medium SNR region with a negligible receiver complexity at the destination.

The rest of this paper is organized as follows. The Section II introduces the proposed NOMA-DRN with the JML. The receivers structures and the benchmark scheme have been also given in this section. Then, in Section III, we define the error performance metrics of the NOMA-DRN with the JML such as bit error rate (BER) and diversity order. We also analyze receiver complexity in this section. Moreover, Section IV present computer simulations to evaluate the error performance of the proposed the NOMA-DRN with the JML along with the comparisons of the benchmark. Finally, Section V presents the conclusion remarks.
\section{System Model}
In this paper, a device-to-device communication is considered where the source ($S$) wants to transmit symbols to the destination ($D$). Due to the large scale objects and/or path-loss, the direct link between $S$-$D$ is not available, hence a cooperative relaying system is implemented with the help of two decode-forward relays (i.e., $R_1$ and $R_2$) which are located between the source and the destination. $R_1$ is to be closer to the destination whereas $R_2$ is closer to the source, thus it is called as diamond relaying. The system model is given in the Fig. 1.  All nodes are assumed to have a single antenna. Since a cooperative communication is needed and the relays operate in half-duplex mode, the end-to-end (e2e) communication is completed in two phases (time slot) where the first phase is between $S-R_i$ (i.e., $i=1,2$) and the second phase is between $R_i-D$. 
\begin{figure}
		\centering
    \includegraphics[width=9cm]{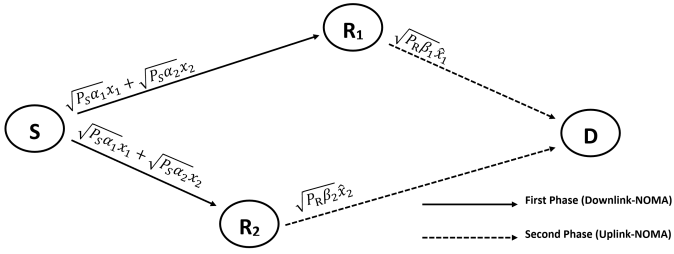}
    \caption{The illustration of the NOMA-based diamond relaying network}
    \label{BER_awgn}
\end{figure}

In order to alleviate the inefficiency of the conventional CRS, NOMA is implemented at the source for two consecutive symbols of the destination. Hence, the two consecutive symbols are superposition coded with the different PA coefficients and broadcasted to the relays in the first phase (downlink-NOMA). The transmitted total symbol is given by
\begin{equation}
x_{sc}=\sqrt{\alpha_1}x_1+\sqrt{\alpha_2}x_2   
\end{equation}
where $\alpha_i, i=1,2$ denotes the PA coefficient for the baseband symbol of the $x_i$ where $\alpha_1>\alpha_2$ is assumed and $\alpha_1+\alpha_2=1$ is satisfied. Hence, the received signal at the relays is

\begin{equation}
y_{R_i}=\sqrt{P_s}\left(\sqrt{\alpha_1}x_1+\sqrt{\alpha_2}x_2\right)h_{SR_i}+n_{R_i}, \ i=1,2
\end{equation}
Where $P_s$ is the transmit power of the source. $h_{SR_i}$ is the channel fading coefficient between $S-R_i$ and the envelopes of it follows Nakagami-m distribution with the $m_{SR_i}$ spread and $\Omega_{SR_i}$ shape parameters. $n_{R_i}$ is the additive Gaussian noise at the receiver $R_i$ and follows $CN(0,N_0)$.

In the second phase, the relays transmit recovered/detected forms of the related symbols to the destination simultaneously. Thus, the second phase can be called as an uplink NOMA. The received signal at the destination is given by
\begin{equation}
y_D=\sqrt{P_{R_1}}\hat{x}_1h_{R_1D}+\sqrt{P_{R_2}}\hat{x}_2h_{R_2D}+n_D
\end{equation}
where $P_{R_i}$ is the transmit power of the relay $R_i$\footnote{The relay powers can be considered as shared among the relays with a PA $\beta$ as being in \cite{Wan2019,9057498,Kara2019} . Besides, the relays can harvest their transmit power from the RF signals in the first phase as in \cite{Karab}. However, energy harvesting is not considered in this paper and it is beyond the scope of this paper.}. $\hat{x}_1$ and $\hat{x}_2$ are the recovered/detected symbols of the $x_1$ and $x_2$ at the relays $R_1$ and $R_2$, respectively. $n_{D}$ is the additive Gaussian noise at the destination and follows $CN(0,N_0)$.

\subsection{Proposed Receiver Structures}
\subsubsection{Decoding at the relays}
According  to (3), the $R_1$ iforwards $x_1$ symbols and $R_2$ transmits $x_2$ symbols. As seen in (1), the $x_1$ symbols has higher PA, hence the $x_1$ symbols can be directly detected at the $R_1$ by pretending $x_2$ symbols as noise. To this end, the maximum likelihood (ML) detector at the $R_1$ is given as
\begin{equation}
    \hat{x}_1=\argmin_{k}{\left|y_{R_1}-\sqrt{P_s\alpha_1}h_{SR_1}x_{1,k}\right|^2}, \ k=1,2,\dots, M_1,
\end{equation}
where $x_{1,k}$ shows the $k$th point in the $M_1$-ary constellation.

On the other hand, since it has lower PA, $x_2$ can not be directly decoded at $R_2$ unlike $x_1$ symbols. An interference mitigation is needed. Thus, we implement an SIC \footnote{We could use joint maximum likelihood (JML) detector as proposed for the second phase rather than the SIC detector. However, the JML and SIC detectors have the same error performance in the downlink NOMA \cite{Al-Dweik2020}. Besides, the JML costs a computational complexity as discussed in Section III. Therefore, the SIC detector is chosen.} where $x_1$ symbols are firstly detected and subtracted from the received signal and then $x_2$ symbols are detected. The detection steps at the $R_2$ are given as
\begin{equation}
    \hat{x}_2=\argmin_{j}{\left|y_{R_2}^{(*)}-\sqrt{P_s\alpha_2}h_{SR_2}x_{2,j}\right|^2}, \ j=1,2,\dots, M_2,
\end{equation}
where
\begin{equation}
    y_{R_2}^{(*)}=y_{R_2}-\sqrt{P_s}h_{SR_2}\sqrt{\alpha_1}\hat{x}_1^{(*)}.
\end{equation}
and
\begin{equation}
    \hat{x}_1^{(*)}=\argmin_{k}{\left|y_{R_2}-\sqrt{P_s\alpha_1}h_{SR_2}x_{1,k}\right|^2}, \ k=1,2,\dots, M_1,
\end{equation}
\subsubsection{Decoding at the destination}
At the destination, both symbols (i.e., $x_1$ and $x_2$) should be detected. Indeed, it can be achieved by a SIC detector likewise at $R_2$ in the first phase. However, unlike the first phase, symbols are exposed to different channel fading coefficients in the second phase (uplink NOMA), and it is known that in the uplink NOMA, the SIC detector does not perform well so that an error floor occurs. Hence, the NOMA-DRN would have a poor error performance. On the other hand, the JML has a better performance in detecting signals for uplink NOMA. To this end, in the second phase, the destination implements a JML detector and it is given as
\begin{equation}
\begin{split}
     [\tilde{x}_1, \tilde{x}_2]=\argmin_{j,k}{\left|y_D-\sqrt{P_{R_1}}h_{R_1D}x_{1,k}-\sqrt{P_{R_2}}h_{R_2D}x_{2,j}\right|^2}, \\
     \ j=1,2,\dots, M_2, \ k=1,2,\dots, M_1,
\end{split}
\end{equation}
where $\tilde{x}_1$, $\tilde{x}_2$ are the detected symbols at the destination of $x_1$ and $x_2$, respectively.

\section{Performance Metrics}
\subsection{Bit Error Rate (BER)}
Since $x_1$ and $x_2$ are two consecutive symbols of the destination, the overall BER performance of the NOMA-DRN can be obtained by averaging BERs of two symbols. It is given by
\begin{equation}
    P_{NOMA-DRN}(e)=\frac{P_1(e)+P_2(e)}{2}
\end{equation}
where $P_1(e)$ and $P_2(e)$ are the e2e BER of the $x_1$ and $x_2$  symbols, respectively. 

The e2e BERs of the symbols are defined as
\begin{equation}
\begin{split}
    P_i(e)&=\frac{1}{M_i}\sum P\left(x_{i}\rightarrow \tilde{x}_{i}\right)   \ i=1,2 \\ &=\frac{1}{M_i}\sum_j\sum_{m\ne j}P\left(x_{i,j}\rightarrow x_{i,m}\right) 
\end{split}
\end{equation}
where $P\left(x_{i}\rightarrow \tilde{x}_{i}\right)$ denotes the pairwise error probability (PEP) when $x_{i}$ is transmitted at the source and detected as $\tilde{x}_{i}$ at the destination. Hence, it is derived by averaging all points in the constellation. Moreover, since the e2e communication consists of two phases and the relay has a DF protocol, the e2e PEP is turns out to be
\begin{equation}
\begin{split}
    &P\left(x_{i}\rightarrow \tilde{x}_{i}\right)= \\
    &P\left(x_{i}\rightarrow \hat{x}_{i}\right)+P\left(\hat{x}_{i}\rightarrow \tilde{x}_{i}\right)-P\left(x_{i}\rightarrow \hat{x}_{i}\right)\cap P\left(\hat{x}_{i}\rightarrow \tilde{x}_{i}\right) 
    \end{split}
\end{equation}
where $P\left(x_{i}\rightarrow \hat{x}_{i}\right)$ and $P\left(\hat{x}_{i}\rightarrow \tilde{x}_{i}\right)$ are the PEPs in the first phase and second phase, respectively. Thus, intersection of them is subtracted to obtain the e2e PEP. Considering this, the e2e BERs of the symbols are given as
\begin{equation}
    P_i(e)=1-\left(1-P_i^{(S-R)}(e)\right)\left(1-P_i^{(R-D)}(e)\right)
\end{equation}
where $ P_i^{(S-R)}(e)$ and $P_i^{(R-D)}(e)$ denote the BER of the first phase and the second phase, respectively. They are defined as given in (10). However, since the transmission strategies differ in two phases (i.e., downlink and uplink NOMA), they are not the same. The $P_i^{(S-R)}(e)$ can be found in \cite[eq.(4)]{Kara2018c} and  \cite[eq.(11)]{Kara2018c} over Nakagami-m fading channels. On the other hand, $ P_i^{(R-D)}(e)$ of JML detector is obtained in \cite[eq.(7)]{Yeom2019} and \cite[eq.(10)]{Yeom2019} for only Rayleigh fading channels. By substituting these equations into (12) and then by substituting (12) into (9), the BER of the NOMA-DRN with the JML is obtained. 
\subsection{Diversity}
The diversity order of the NOMA-DRN is given by
\begin{equation}
    \nu=\lim_{SNR\rightarrow\infty}\frac{\log P_{NOMA-DRN}}{\log SNR}
\end{equation}

Since the  $P_{NOMA-DRN}$ is the average of $P_1(e)$ and $P_2(e)$, the diversity order of the NOMA-DRN is given by
\begin{equation}
    \nu=\min\{\nu_1,\nu_2\}
\end{equation}
where $\nu_1$ and $\nu_2$ are the diversity orders of the symbols $x_1$ and $x_2$, respectively. Moreover, since a cooperative communication is included, the diversity order of the symbols is limited by the weakest link. Thus, it is given that
\begin{equation}
    \nu_i=\min\{\nu_i^{(S-R)},\nu_i^{(R-D)}\}
\end{equation}
where $\nu_i^{(S-R)}$ and $\nu_i^{(R-D)}$ are given the diversity orders of the $x_i$ symbols in the first phase (downlink NOMA) and in the second phase (uplink NOMA), respectively. The diversity order of $x_i$ in the downlink NOMA is given as $m_{SR_i}$ in \cite{Kara2018c}. On the other hand, the diversity order of the uplink NOMA with the JML is given for Rayleigh fading channels as the number of receiving antennas (i.e., $1$ in this paper.) With the extension to the Nakagami-m fading channels, it can be said as the $m_{R_iD}$. However, this is only valid in the low-medium SNR region. With the increase of the SNR, the difference between the channel qualities/transmit powers becomes dominant on the diversity order rather than the shape parameters and the diversity order of the uplink NOMA in the high SNR region is limited by 1. To this end,  by substituting these into (15) and then into (14), the diversity order of the NOMA-DRN with the JML  in the low-medium SNR region is obtained as
\begin{equation}
    \nu=\min\{m_{SR_1},m_{SR_2},m_{R_1D},m_{R_2D}\}.
\end{equation}
whereas it is limited by $\nu=1$ in the high SNR region. 

\subsection{Complexity}
In this section, we provide a comparison between the SIC based NOMA-DRN and the proposed NOMA-DRN with the JML in terms of the receiver complexity at the destination. The receivers complexities of the relays are not considered since they are the same in both the  SIC based NOMA-DRN and the proposed NOMA-DRN with the JML. For the complexity compassion, we derive the number of the required complex operations. To obtain that, according to (5)-(7) (it is similar for the SIC detector at the destination), the SIC detector firstly generates $M_1$ candidates for the detection of the $x_1$ symbols, then it calculates the Euclidean distances and choose the minimum one by comparing. After detecting the $x_1$ symbols, it subtracts its regenerated form from the received signal. Finally, the above steps are repeated for the detection of $x_2$ symbols with $M_2$ candidates. Therefore, by considering $M_1=M_2=M$ (for notation simplicity), the required complex operations is given as
\begin{equation}
    \delta_{SIC}=\underbrace{2M+1}_{Adder}+\underbrace{4M}_{Multiplier}+\underbrace{2(M-1)}_{Comparator}=8M-1
\end{equation}

On the other hand, in the JML detector as given in (8), the joint $M_1\times M_2$ candidates are generated and the Euclidean distances are computed for each. Then, the minimum is obtained within these distances. To this end, the complexity of the JML detector is obtained as
\begin{equation}
    \delta_{JML}=\underbrace{2M^2}_{Adder}+\underbrace{3M^2}_{Multiplier}+\underbrace{M^2-1}_{Comparator}=6M^2-1.
\end{equation}

In order to compare the complexities of both receivers, we present the total number of complex operations for different modulation sizes in Table I. As one can easily see that the complexity of the JML is negligible when M is relatively low. On the other hand, it increases exponentially with the increase of M. However, by considering the performance gain (as discussed in the next section), this complexity increase is affordable. Besides, because of this complexity, as we discussed in the Section 2, we implement SIC based detectors at the relay since both the SIC and JML detectors have the same error performance in downlink NOMA.  

\begin{table*}[t]
\centering
\caption{Complexity Comparisons Between SIC and JML detectors}
\begin{tabular}{c||c|c|c|c||c|c|c|c} \hline
&\multicolumn{4}{c||}{SIC detector} &\multicolumn{4}{c}{JML detector} \\ \hline
$M$&Adder&Multiplier&Comparator&Total&Adder&Multiplier&Comparator&Total\\  \hline \hline
2&5&8&2&15&8&12&3&23 \\ \hline
4&9&16&6&31&32&48&15&95 \\ \hline
8&17&32&14&63&128&192&63&383 \\ \hline
16&33&64&30&127&512&768&255&1535\\ \hline 
\end{tabular}
\label{table3}
\end{table*}
\section{Numerical Results}
In this section, we present computer simulations to evaluate the BER performance of the proposed NOMA-DRN with the JML. Besides, we also provide comparisons with the conventional NOMA-DRN with the SIC detectors. For fair comparisons with \cite{Kara2019}, in the simulations, we consider a total transmit power for the relays (i.e., $P_R$) and it is shared among the relays as $P_{R_1}=\beta P_R$ and $P_{R_2}=(1-\beta)P_R$. In simulations, $P_s=P_R$ is assumed Firstly, we present BER simulations when $m_{SR_1}= m_{SR_2}= m_{R_1D}= m_{R_2D}=1$ (Rayleigh fading) in Fig. 2 to compare the results with \cite{Kara2019}. In Fig. 2, the shape parameters are $\Omega_{SR_1}^2=2$, $\Omega_{SR_2}^2=10$, $\Omega_{R_1D}^2=9$, $\Omega_{R_2D}^2=3$.  The PA coefficients are assumed to be $\alpha=0.9602$ and $\beta=0.8011$ as the same with \cite{Kara2019} and noted as the optimum values in \cite{Wan2019}. The modulation orders are selected as BPSK for both symbols.  In the simulations, we give results for the BERs of both symbols and for averaged NOMA-DRN. As seen in Fig. 2, with the proposed JML in the second phase, the error performance of the NOMA-DRN has been improved significantly and an error floor does not exist any more. 
\begin{figure}
		\centering
    \includegraphics[width=9cm]{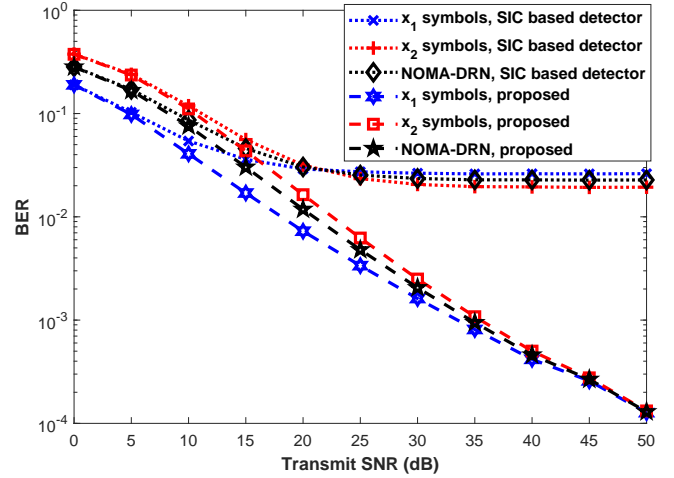}
    \caption{BER Comparisons of the NOMA-DRN with SIC-based detector and the proposed JML detector over Rayleigh fading channels when  $\Omega_{SR_1}^2=2$, $\Omega_{SR_2}^2=10$, $\Omega_{R_1D}^2=9$, $\Omega_{R_2D}^2=3$, $\alpha=0.9602$ and $\beta=0.8011$ }
    \label{BER_rayleighn}
\end{figure}
\begin{figure}
		\centering
    \includegraphics[width=9cm]{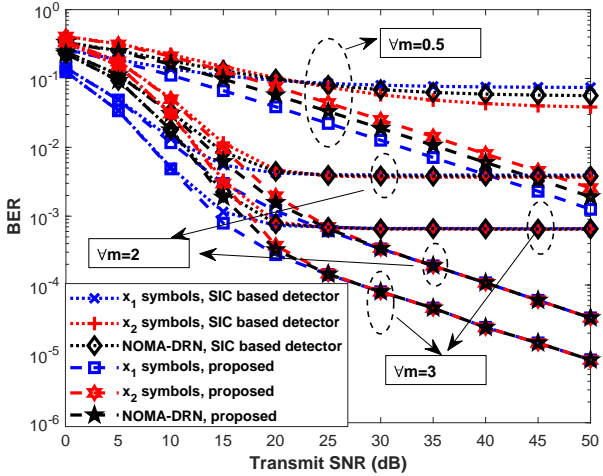}
    \caption{BER Comparisons of the NOMA-DRN with SIC-based detector and the proposed JML detector over Nakagami-m fading channels with different shape parameters when all shape parameters are the same ($\forall m=0.5,2,3$)  $\Omega_{SR_1}^2=2$, $\Omega_{SR_2}^2=10$, $\Omega_{R_1D}^2=9$, $\Omega_{R_2D}^2=3$, $\alpha=0.9602$ and $\beta=0.8011$ }
    \label{BER_nakag_m}
\end{figure}

To reveal the effects of the shape parameters on the diversity order, we present simulation results over Nakagami-m fading channels for  $m_{SR_1}= m_{SR_2}= m_{R_1D}= m_{R_2D}=0.5$, $m_{SR_1}= m_{SR_2}= m_{R_1D}= m_{R_2D}=2$ and $m_{SR_1}= m_{SR_2}= m_{R_1D}= m_{R_2D}=3$ in Fig. 3. The spread parameters are the same with Fig. 2.  As seen in Fig. 3, the diversity order of the NOMA-DRN with JML is equal to the shape parameters (i.e., $0.5,2$ and $3$) for all scenarios in the low-medium SNR region. On the other hand, in the high SNR region it is limited by $1$ when the shape parameter is greater than $1$. Nevertheless, with the increase of shape parameters, a horizontal performance gain is achieved although the diversity order is not changed. For instance, the NOMA-DRN with shape parameters $3$ has $\sim 10\ dB$ better performance than the scenario with the shape parameters of $2$. Furthermore, regardless of the channel conditions, the proposed NOMA-DRN with JML outperfoms the NOMA-DRN with the SIC detectors significantly. The proposed NOMA-DRN has no error floor and the performance gain over the SIC detector is up to $\sim 20-25 dB$ in some scenarios.   
\begin{table*}[t]
\centering
\caption{Shape and Spread Parameters in Fig. 4 and Fig. 5}
\begin{tabular}{c||c|c|c|c|c|c|c|c} \hline
Scenario& $m_{SR_1}$&$\Omega_{SR_1}$&$m_{SR_2}$&$\Omega_{SR_2}$&$m_{R_1D}$&$\Omega_{R_1D}$&$m_{R_2D}$&$\Omega_{R_2D}$ \\ \hline \hline
I&3&1&3&10&1&10&1&1 \\ \hline
II&3&5&3&10&1&10&1&5 \\ \hline
III&1&1&1&10&3&10&3&1 \\ \hline
\end{tabular}
\label{table}
\end{table*}

Moreover, to show the effect of the minimum shape parameter and the spread parameters, in Fig. 4, we present the error performance of the NOMA-DRN with the JML for different scenarios. The parameters in each scenario are given in Table II. As seen in Fig, 4, all scenarios have the diversity order of the 1 since the minimum shape parameters between nodes is equal to 1. Nevertheless, according to the spread parameters, the error performance of the system changes. For instance, between Scenarios I and II, the error performance of the $x_1$ symbols gets worse since the channel qualities difference between $R_1-D$ and $R_2$-D becomes less so that the error performance of the second phase (uplink NOMA) is decreased. Hence, this decrease dominates the e2e performance of the $x_1$ symbols although the $\Omega_{SR_1}$ is increased. On the other, the second phase performance (uplink NOMA) of the $x_2$ symbols is increased with the increase of the $\Omega_{SR_1}$. Thus, the overall performance of the NOMA-DRN in Scenario II is better than the Scenario I. Compared the Scenario I and III, although the both scenarios have the same diversity order, the Scenario III has better error performance. This explained as follows. In scenario III, with the increase of the shape parameters between the relays and the destination ($1\rightarrow3$), the error performance of the second phase (uplink NOMA) is improved and the second phase of the communication limits the error performance of the NOMA-DRN. Thus, this improvement provides a gain in overall performance of the NOMA-DRN. To validate above discussions, we provide the error performances of the both phases in Fig. 5. One can easily see that, although it is increased by the JML detector, the second phase (uplink NOMA) has worse performance than the first phase (downlink NOMA). Therefore, when the second phase performance is better, the NOMA-DRN has a good error performance. 
\begin{figure}
		\centering
    \includegraphics[width=9cm]{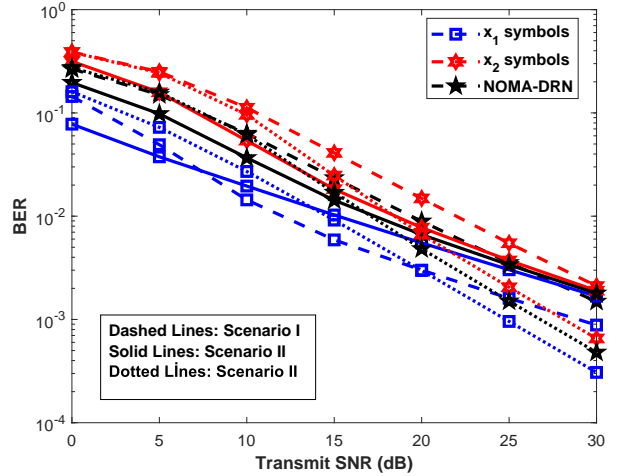}
    \caption{BER of the NOMA-DRN with the proposed JML detector over Nakagami-m fading channels for the scenarios given in Table II }
    \label{BER_nakag_m_diff}
\end{figure}
\begin{figure}
		\centering
    \includegraphics[width=9cm]{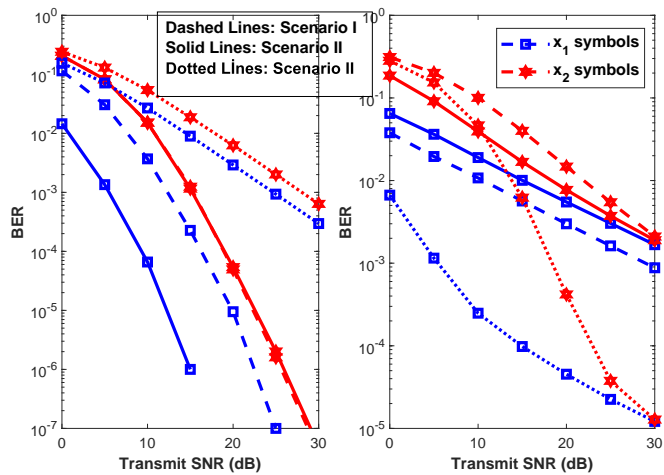}
    \caption{BER of the both phases in the NOMA-DRN with the JML for the scenarios given in Table II a) First phase (downlink NOMA) b) Second phase (uplink NOMA) }
    \label{BER_nakag_m_dl_ul}
\end{figure}

\ifCLASSOPTIONcaptionsoff
  \newpage
\fi



%
\section{Conclusion}
In this paper, we propose a novel NOMA-DRN with the JML at the destination. We demonstrate the error performance of the proposed network and proved that the NOMA-DRN has  not an error floor unlike the benchmark (NOMA-DRN with the SIC detector.) It outperforms remarkably the benchmark and a reliable communication can be accomplished with the increased spectral efficiency compared to the conventional CRS schemes. Furthermore, we also provide diversity analysis and the complexity analysis for the proposed network. We present that the proposed network can achieve full diversity order in the low-medium SNR region and the receiver complexity is affordable by considering the error performance gain. Moreover, the proposed model can save transmit energy up to $\sim 20-30dB$ compared to the benchmark. This is very promising for the energy-constrained networks such as IoT.   

\bibliographystyle{IEEEtran}
\bibliography{bibtex}

%






\end{document}